\documentclass[twocolumn,superscriptaddress,preprintnumbers,amsmath,amssymb,floatfix]{revtex4}
\usepackage{epsfig}
\usepackage{epstopdf}
\usepackage{graphicx}
\usepackage{grffile}
\usepackage{bm}
\usepackage{sidecap}
\usepackage{mathptmx}
\usepackage{txfonts}
\usepackage[colorlinks,citecolor=blue,linkcolor=blue,urlcolor=blue]{hyperref}
\usepackage[usenames,dvipsnames,svgnames]{xcolor}
\usepackage{array}
\usepackage{float}
\usepackage{lipsum}
\begin{document}
\title{Impacts of in-plane strain on commensurate graphene/hexagonal boron nitride superlattices}
\date{\today}

\author{Zahra Khatibi}\email{za.khatibi@gmail.com}
\affiliation{Department of Physics, Iran University of Science and Technology, Narmak, 16846-13114, Tehran, Iran}
\affiliation{Chalmers University of Technology, Department of Physics, 412 96 Gothenburg, Sweden}
\author{Afshin Namiranian\footnote{Author to whom any correspondence should be addressed}}
\affiliation{Department of Physics, Iran University of Science and Technology, Narmak, 16846-13114, Tehran, Iran}
\author{S. F. K. S. Panahi}
\affiliation{Department of Physics, Iran University of Science and Technology, Narmak, 16846-13114, Tehran, Iran}

\begin{abstract}
Due to atomically thin structure, graphene/hexagonal boron nitride (G/hBN) heterostructures are intensively sensitive to the external mechanical forces and deformations being applied to their lattice structure. In particular, strain can lead to the modification of the electronic properties of G/hBN. Furthermore, moir\'e structures driven by misalignment of graphene and hBN layers introduce new features to the electronic behavior of G/hBN. Utilizing {\it ab initio} calculation, we study the strain-induced modification of the electronic properties of diverse stacking faults of G/hBN when applying in-plane strain on both layers, simultaneously. We observe that the interplay of few percent magnitude in-plane strain and moir\'e pattern in the experimentally applicable systems leads to considerable valley drifts, band gap modulation and enhancement of the substrate-induced Fermi velocity renormalization. Furthermore, we find that regardless of the strain alignment, the zigzag direction becomes more efficient for electronic transport, when applying in-plane non-equibiaxial strains. 
\end{abstract}
 \maketitle

\section{Introduction}\label{intro}
Few-layered heterostructures constructed of 2D materials, consisting of graphene and its isomorphs, such as hexagonal boron nitride, transition metal dichalcogenides, etc., are introduced as an alternative to graphene for band gap emerging and their versatile and prosperous properties \cite{Georgiou2012,Lee2013, Roy2013,Shayeganfar2016}. Since different 2D layers have diverse elastic and electronic properties, the final properties of heterostructures are strongly affected by strain and stacking, and thus can be tuned to fit new functionalities \cite{Novoselov2012}. Graphene/hexagonal boron nitride (G/hBN) has surpassed other heterostructures in research since it can be employed as an approach to control the electronic properties of graphene leaving its favorite features like mobility unchanged \cite{Gannett2011, Xue2011,Tang2015}. This is due to the fact that hBN as a substrate possesses flat surface and imposes less charge inhomogeneity in graphene \cite{Dean2010}. G/hBN heterostructures also have demonstrated signatures of tunability of band gap regarding the application of strain, corrugation and misalignment of layers \cite{Bokdam2014, San-Jose2014,Cosma2014,Wallbank2015}. To date, different band gap magnitudes for G/hBN heterostructures have been reported by both theoretical and experimental studies \cite{Giovannetti2007,Sachs2011,Song2013,Amet2013,Hunt2013,Wang2016}. The $50~\rm{meV}$ band gap predicted by early {\it ab initio} study \cite{Giovannetti2007} of band structure of commensurate G/hBN was objected by recent theoretical studies \cite{Sachs2011,Xue2011}. The reason was argued to be the intrinsic strains due to ignorance of the lattice mismatch between graphene and the underlying hBN. In fact, the incommensurability effects were shown to be responsible for the cancellation of quasiparticle band gaps in more realistic systems with the inclusion of lattice mismatch \cite{Bokdam2014,Yankowitz2016a}. Yet, interestingly, a nearly $160~\rm{meV}$ band gap at the primary graphene Dirac cone for $0^{\circ}$-aligned G/hBN has been observed, confirming the fact that the physics of G/hBN is not completely known \cite{Wang2016}. It is believed firstly, that the nature of the method of measurements itself, and secondly, the increment of the mass term as a consequence of substantial height variation, in-plane strain, and reduction of interlayer distance, are the main reasons for the observation of a large band gap compared to previous studies. 

Rotationally misaligned neighboring layers, alongside the lattice mismatch of crystallographic structure between graphene and hBN leading to quasi-periodic structures, moir\'e pattern \cite{Yankowitz2012,Woods2014}, has also been studied extensively for their interesting electronic properties \cite{Mishchenko2014,Kvashnin2015,Guisset2016,Jung2017}. The modulations include renormalization of Fermi velocity \cite{Chizhova2014}, fractal quantum Hall effect \cite{Dean2013,Hunt2013}, the emergence of Secondary Dirac cones and band gaps \cite{Ponomarenko2013,Wang2016}. Also, commensurate-incommensurate transition in moir\'e patterns as a result of lattice adjustment of graphene to its substrate is shown, both theoretically and experimentally \cite{Woods2014,Jung2015}. These transitions can accumulate strains leading to modification of the electronic properties of G/hBN\cite{Woods2014}.

Despite the fact that the G/hBN heterostructures have been extensively investigated, yet the possible enhancement of the impacts of the external strain on the electronic properties of G/hBN by the moir\'e superstructures is a less-discussed intriguing open question. In this paper, using {\it ab initio} calculation, we study the consequences of in-plane strain on commensurate G/hBN with large misalignment angle, when the strain is applied to both layers simultaneously. The main idea is to study the use of moir\'e pattern in van der Waals heterostructures for magnification of strain effects, such as modification of the gap energy, especially when the twist angle of the layers becomes large and the superlattice periodicity is typically small (e.g. $2~\rm{nm}$). We show that the interplay of the few percent magnitude homogeneous strains and moir\'e pattern in the experimentally applicable system presents large band gap tunability. This feature can be exploited in 2D material-based nanoelectronic devices. Devices in which, the maintenance and successful fabrication of heterostructures with a controlled rotational angle of layers and desired in-plane strain, is shown to be experimentally feasible \cite{Kim2016,Yankowitz2016a}.

The paper is structured as follows.
In Sec.\ref{theo}, we address the details of the geometrical structure of commensurate G/hBN superlattices and the DFT calculation. 
In Sec.\ref{res}, we present the results for the strain energy, band dispersion, gap energy, and band velocity for unstrained and strained G/hBN heterostructures. We conclude our findings in Sec.\ref{concl}.
\section{Theory}\label{theo}
In this section, we briefly address the geometric definition of the commensurate moir\'e structures. A detailed discussion on the derivation of the commensurate moir\'e superlattice vectors and the angle at which these structures occur, can be found in Ref.
\cite{Lopes2012} and Ref. \cite{Mele2012}. We, then, introduce and calculate the electronic properties of three different commensurate structures, both before and after applying strain.

For two layers of honeycomb lattices that are rotated with respect to each other around the normal vector to their planes, commensurate structures take place in discrete angles $\theta_{mn}$. These angles are those at which the lattice translation vectors of the upper and lower layers, $\vec{T}_{mn}=m\vec{a}_1+n\vec{a}_2$ and $\vec{T}^{\prime}_{mn}=m^\prime \vec{a}^{\prime}_1+n^\prime \vec{a}^{\prime}_2$, on the span of their primitive vectors, $a_{1(2)}~,~a^{\prime}_{1(2)}$, become equal, i.e. $\vec{T}_{mn}=\vec{T}^\prime_{m^\prime n^\prime}$. Here, $m$ and $n$ are integers and $\vec{T}_{mn}$ denotes the position of the $A$ sublattice in $(m,n)$ cell of upper layer whereas $\vec{T}^\prime_{m^\prime n^\prime}$ is the $A$ sublattice vector in lower layer. Therefore, the commensuration condition yields that starting from fixed $A$ sublattices of both layers at the origin, the next $A$ sublattice of each layer meet when $\vec{T}_{mn}=\vec{T}^\prime_{m^\prime n^\prime}$. Accordingly, it can be shown that, the total number of disclosed atoms in commensurate supercell is $4(n^2+nm+m^2)$, and the relative misalignment angle is $\cos^{-1}\left({\frac{n^2+4nm+m^2}{2(n^2+nm+m^2)}}\right)$ \cite{Uchida2014}. 

\begin{figure}[!htbp]
\centering
\includegraphics[width= .6\linewidth]{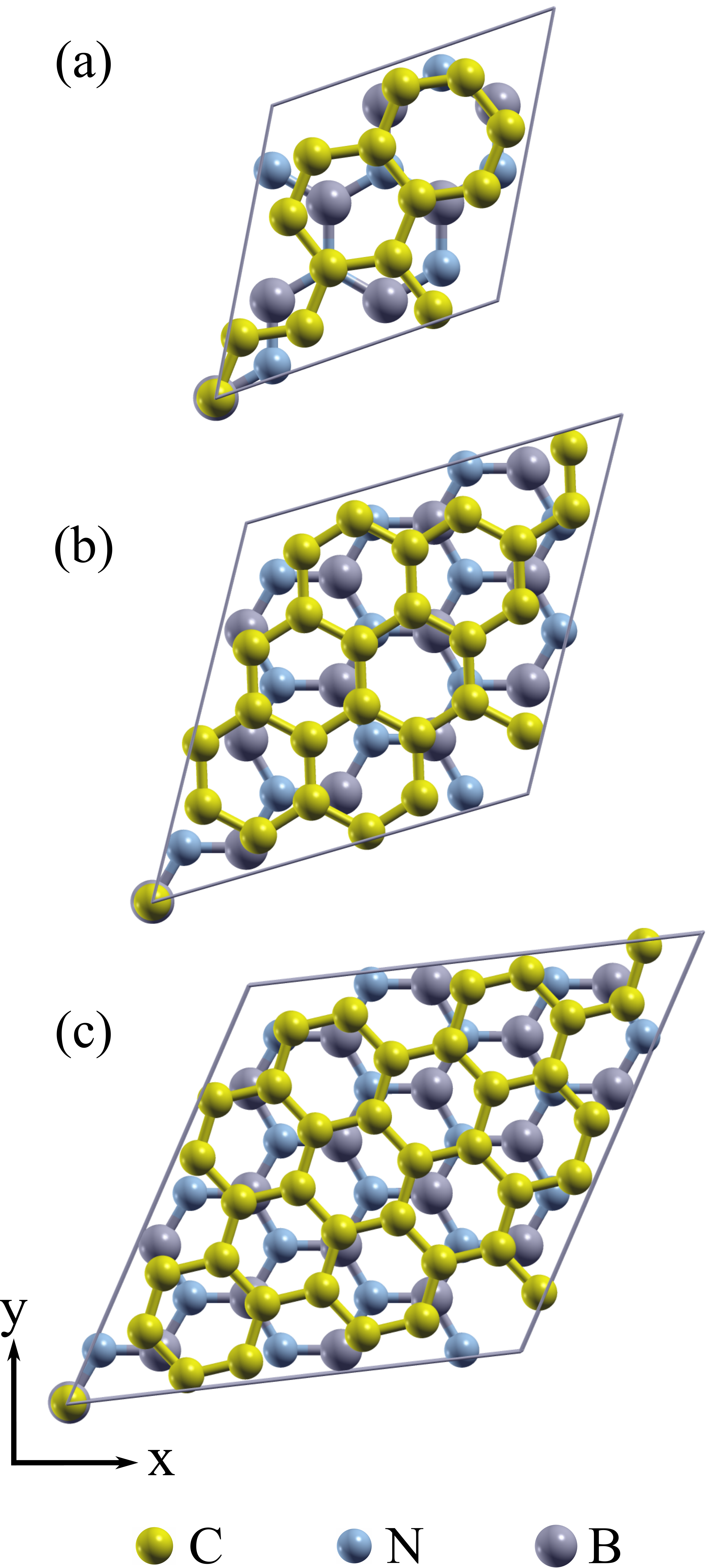}
\caption{(Color online). 
Top view of the schematic representation of (a) $\alpha$ (b) $\beta$ and (c) $\lambda$ commensurate supercells of single layer graphene on hBN with the misalignment angle of $\theta \approx 38.21 ^{\circ}$, $\theta \approx 32.20^{\circ}$ and $\theta \approx 13.17 ^{\circ}$, respectively. \label{fig1}}
\end{figure}

In this article, we study three non-equivalent $(m,n)$ commensurate structures (See Fig.\ref{fig1}), listed below \footnote{the 1.8\% lattice mismatch between the graphene and hBN layer is been disregarded to ease the simulation of G/hBN according to the current scope of the possibility of {\it ab initio} calculations \cite{Wijk2014}}:
\begin{itemize}
\item[]
($\alpha$) commensurate supercell $(1,4)$ with a misalignment angle of $38.21 ^{\circ}$ containing $28$ atoms in total.
\item[]
($\beta$) commensurate supercell $(1,3)$ with a twist angle of $32.20^{\circ}$ and a total number of $52$ atoms in the supercell.
\item[]
($\lambda$) commensurate supercell $(2,3)$ with a misalignment angle of $13.17 ^{\circ}$ constituted of $76$ atoms in total.
\end{itemize}
Also, for the sake of convenience and simplicity, we name the G/hBN superstructures $\alpha$, $\beta$ and $\lambda$, henceforth throughout this study, respectively.

%
%
%
%
%
%
We study the electronic behavior of these G/hBN structures performing first-principles calculations implemented in the SIESTA code \cite{Soler2002}. Double-$\zeta$ polarized basis (DZP), with Norm-conserving pseudopotential, alongside the vdW-DF exchange-correlation functional (DRSLL \cite{Dion2004}), within the conjugate gradient method is employed. For each structure demonstrated in Fig.\ref{fig1}, the lattice constant, k-grid and mesh cut-off convergence tests have been done, to ensure the consumption of best sets of inputs while same optimized basis is used for relaxation procedure. In this regard, $10\times 10 \times 1$ Monkhorst-Pack k-point grids and $400~{\rm Ry}$ mesh cutoff are used for moir\'e structures $\alpha$, $\beta$ and $\lambda$, respectively. All simulations include a vacuum space of approximately 20~\AA~height to exclude any interactions between spurious images of the G/hBN. For an unstrained system, both atomic coordinates and lattice vectors are allowed to relax so that the forces on each atom become less than 0.04 eV/\AA.

Strain can affect the electronic properties of G/hBN through the distortion of the lattice structure which leads to the modification of the overlaps of the atomic orbitals. While uniaxial strain distorts the lattice anisotropically, the biaxial strain restores the lattice symmetries and expands the lattice homogeneously along the in-plane axes. Moreover, when applying compressive strain, atoms are pushed closer to each other and in extreme cases where the strain is large, corrugations may occur. Here, we study the impacts of in-plane strains in both biaxially strained systems and mixed situations where the G/hBN structure is uniaxially strained in one direction and compressed along the other direction. To strain moir\'e structures within the DFT method, we model the lattice vectors for each of the superlattices depicted in Fig.\ref{fig1} as $\vec{R}^{\prime}_{i}=(1+\varepsilon_{i})\vec{R}_{i}$, in which $\varepsilon_{i}$ is the strain component along the in-plane direction, $i=x,y$. Next, we optimize the lattice structure while the lattice vectors are set to be fixed at their strained values and only the atomic coordinates are allowed to relax in accordance with the vectors.
\section{Results and discussions}\label{res}
In low energy regime, where the most striking electronic properties of graphene emerge, the electronic behavior of the G/hBN is mostly driven from that of graphene since hBN is a large band gap insulator \cite{Watanabe2004,pacile2008}. Furthermore, being an atomically thin zero-gap semiconductor, graphene is sensitive to the external mechanical forces and can be strongly affected by in-plane strain being exerted on its lattice structure \cite{GUINEA2012,Choi2010a,AMORIM2016}. Therefore, the electronic properties of G/hBN are modified by strain mainly through geometrical changes and modulation of the inter-atomic distances leading to changes of the overlaps of the atomic orbitals. Also, strain affects the interlayer interactions of G/hBN through expansion and compression of the lattice structure of the individual layers. In the following section, we calculate the strain energy and discuss the G/hBN electronic band dispersion near the charge neutrality point before and after application of strain. Next, we address gap modulation of G/hBN moir\'e patterns in accordance with biaxial and non-equibiaxial strains.

Our DFT computation shows that the mean value of the relaxed interlayer distance between graphene and hBN for $\alpha$, $\beta$ and $\lambda$ is 3.5096 ~\AA, 3.4425~\AA, and 3.4573~\AA, severally. The difference between the interlayer distances can be attributed to the misalignment angle between the adjacent layers. In fact, the diverse stacking configurations result in different long- and short-range interactions between the layers and the vertical relaxation strains of G/hBN \cite{Jung2015}. Also, our DFT results are in good agreement with previous studies \cite{Giovannetti2007,Kan2012,Bokdam2014}. Among symmetric stackings of G/hBN (AB, BA, and AA), the AB stacking, i.e. a Nitrogen atom located on top of the center of a graphene hexagon has smaller on-site energy deviations from that of graphene, and is more energetically favorable \cite{Wijk2014, Jung2014}. $\beta$ superlattice is the most resembling structure to AB-stacked G/hBN since its rotation angle deviates only 2$^\circ$ from that of AB-stacked G/hBN and hence possesses the largest number of Nitrogen atoms located on top of the center of graphene hexagon per supercell. Therefore, it has the smallest interlayer distance among the three.
%
%
%
%
%
%
%
\begin{figure}[!htbp]
\centering
\includegraphics[width=\linewidth]{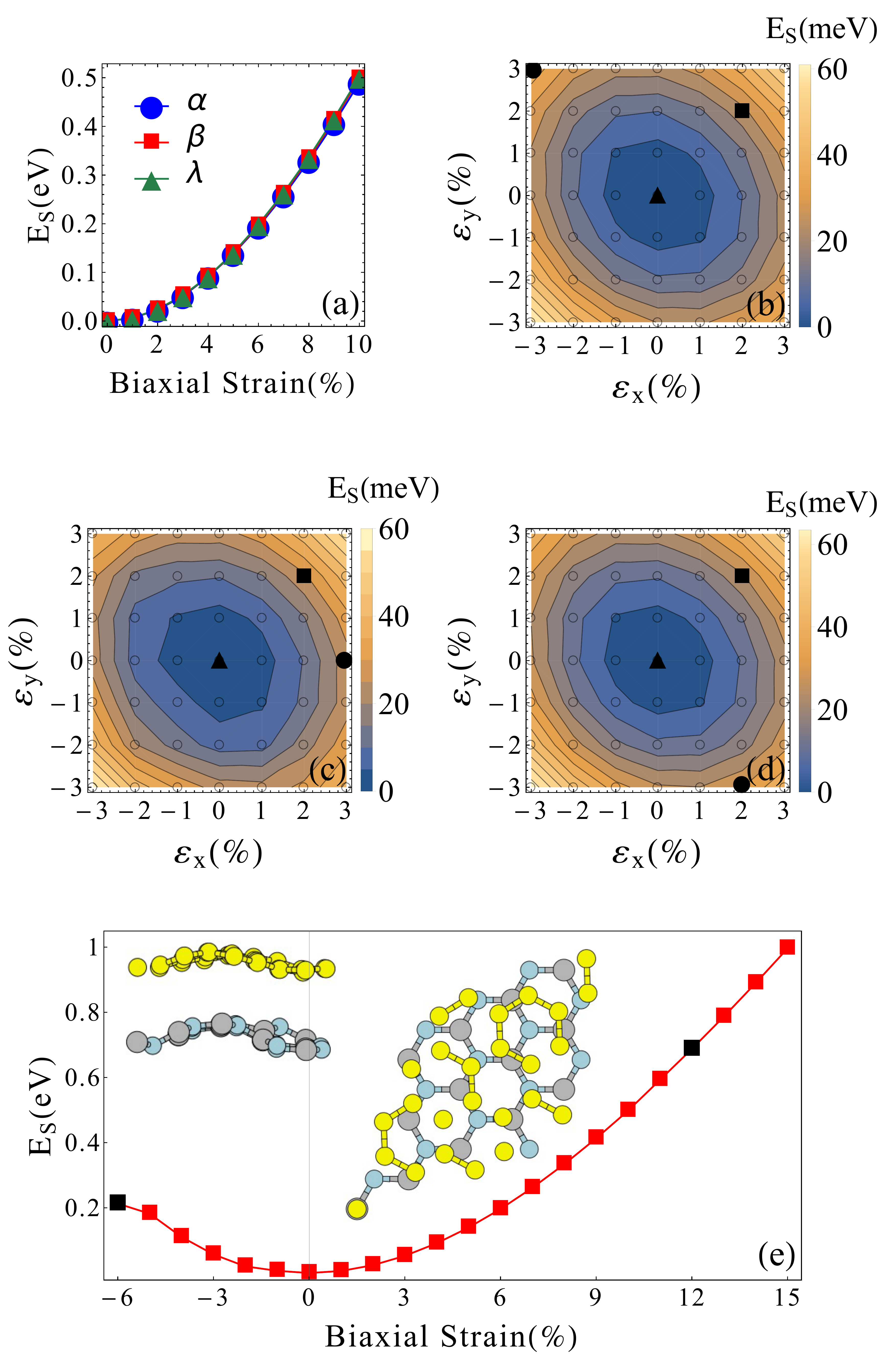}
\caption{(Color online). 
(a) Per-supercell strain energy versus biaxial strain for diverse commensurate moir\'e structures shown in Fig.\ref{fig1}. (b-d) Strain energy maps as a function of in-plane strain strength, for $\alpha$, $\beta$ and $\lambda$ superlattices, respectively. 
The empty small circles correspond to real data and the background is a second-order polynomial which is fitted to the real data. The three black symbols in surface plots are the specific strain configuration that will be used later for band representation in Fig.\ref{fig3}. (e) Strain energy as a function of biaxial strain for $\beta$ in an extended range of applied strain values. The insets are the side view of corrugated G/hBN when being subjected to $6\%$ compressive strain (left) and the top view of the disintegrated structure of G/hBN when $12\%$ biaxial strain is applied (right). \label{fig2}}
\end{figure}

Strain energy is defined as the change in total energy of the system after being deformed through the application of strain, i.e. $E_s=E(\varepsilon)-E^{eq}$ \cite{Averill2011,Peng2012,Wei2014,Elahi2015}. To evaluate the strain energy for each strain configuration we first optimize both strained and unstrained G/hBN lattice structures. As the lattice optimization computations give access to the total energy of the system, we then can obtain per-supercell strain energy via calculating the difference between the total energies of the strained and unstrained G/hBN superstructures and dividing the result by the number of atoms in the supercell. Per-supercell strain energy of G/hBN as a function of applied strain for three commensurate superlattices mentioned earlier is illustrated in Fig.\ref{fig2}. All three superlattices yield the same results and the per-supercell strain energy ranges from zero to $0.5~{\rm eV}$ for $0-10\%$ biaxial strain (See Fig.\ref{fig2}(a)). Also, the strain energy demonstrates no dependence on the misalignment angle, as all three superlattices show the same trend and similar strain energy for all biaxial strain values. As in the case of biaxial strain, the surface plot of per-supercell strain energy of G/hBN displays almost equal response to strain imposition in all superlattices where a rise in strain energy to $60~\rm{meV}$ occurs when strain approaches $3\%$ (cf. Fig.\ref{fig2}(b-d)). Moreover, as it can be seen from surface plots of per-supercell strain energy, non-equibiaxial strains in which the structure is stretched in one direction and compressed in opposite direction are less efficient in the modulation of the total energy of the system compared to biaxial (compressive) strains.

Fig.\ref{fig2}(e) illustrates the per-supercell strain energy for biaxial and compressive strain being applied on $\beta$ G/hBN for a larger range of strain magnitude. Right (left) inset is the lattice structure of the G/hBN supercell when 12\% biaxial (6\% compressive) strain is applied. Comparing the effects of biaxial and compressive strain, we observe that the per-supercell strain energy is approximately equal for both strains (cf. per-supercell strain energy for $\pm$6\% in Fig.\ref{fig2}(e)). Furthermore, while G/hBN experiences severe changes in construction when being subjected to 12\% biaxial strain (right inset of panel (e)), the per-supercell strain energy doesn't show any signs of entering plasticity region which is expected to occur as in the case of other 2D materials \cite{Topsakal2009,Tianshu2012,Zhang2013b,Elahi2015,Wei2014}. This can be ascribed to the fact that the system under study is not constituted purely of either graphene or hBN. The competition between interlayer interaction to maintain G/hBN structure and the interatomic interaction in graphene to overcome the strain effects imposes great impacts on G/hBN structure. Hence, graphene exhibits no structure maintenance for the strains greater than $12\%$ and decomposes. In contrast, hBN is still stretchable and can endure greater strains so that its total energy still increases in accordance with the applied strain \cite{Peng2012}. Therefore, we observe an increment in per-supercell strain energy in general, while the system has already been disintegrated. 

The compressive strain, on the other hand, imposes a totally opposite effect on the G/hBN. The compression also does enlarge the per-supercell strain energy as one gradually go beyond the relaxed structure of G/hBN. The exception is that hBN, due to the larger lattice constant and hence having much-extended structure when isolated, resists large contractions. Consequently, G/hBN starts to corrugate when being exposed to $6\%$ biaxial strain (cf. the left inset of Fig.\ref{fig2}(e)).

%
%
%
%
%
\begin{figure*}[!htbp] 
\centering
\includegraphics[width=.9\linewidth]{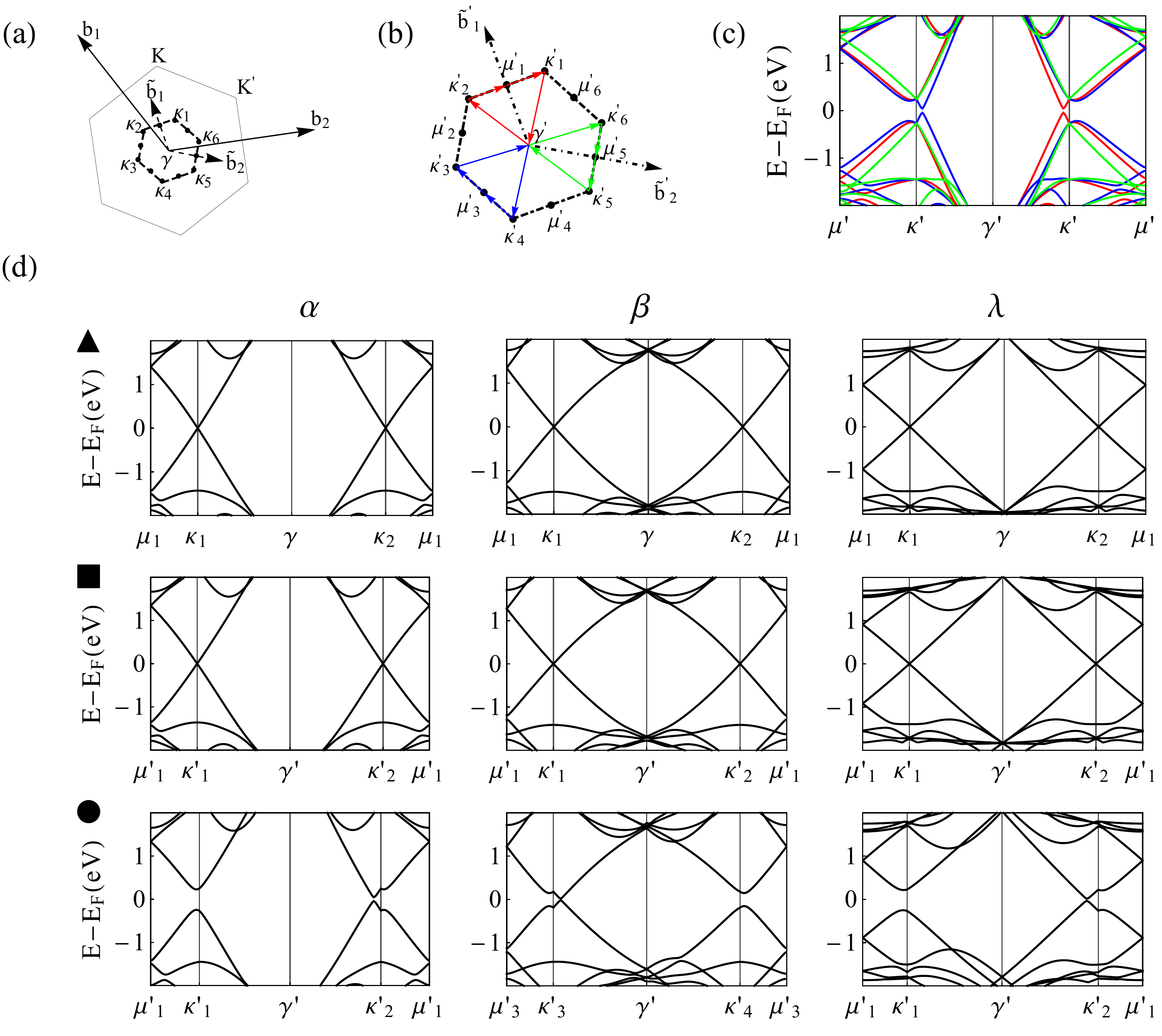}
\caption{(Color online). (a) BZ of graphene and G/hBN. Solid (dashed) hexagon is BZ of graphene (sBZ of G/hBN with $\alpha$ superlattice). $\bf{b}_{1(2)}$ and $\tilde{\bf{b}}_{1(2)}$ are reciprocal lattice vectors of graphene and G/hBN respectively. 
(b) The sBZ for the strained ($\varepsilon_x=-3\%$ and $\varepsilon_y=3\%$) $\alpha$ superlattice of G/hBN. The colored arrows indicate the paths along which, the bands are calculated and presented in panel (c). The G/hBN sBZ is magnified for better clarification of the paths. (c) band dispersion of $\alpha$ G/hBN along the paths illustrated in panel (b). (d) is the band structure of $\alpha$, $\beta$ and $\lambda$ moir\'e superlattice where G/hBN is unstrained (triangle), is subjected to biaxial strain (rectangle) and strained differently along the in-plane axes (circle). The strain configurations for which the band dispersions are brought in panel (d) are denoted by the corresponding symbols in Fig.\ref{fig2}. Here we only demonstrate the paths along which the smallest direct gap occurs. 
\label{fig3}}
\end{figure*}
In Fig.\ref{fig3}, we study the modifications of electronic properties due to strain and plot the band dispersion near the charge neutrality point. Prior to changes of the band gap magnitude, we first comprehend the band topology and relocation of the band gap. As discussed earlier, deformation, both biaxial and non-equibiaxial, distorts lattice structure which leads to modification of the Brillouin zone (BZ). Therefore, the supercell Brillouin zone (sBZ) corners of the unstrained system do not coincide with those of the system under tension. Also, sBZ corners named here as $\kappa^\prime$ points, are no longer equivalent for non-equibiaxially strained G/hBN. Due to time reversal symmetry, there are two groups of equivalent sBZ corners. Hence, as illustrated in Fig.\ref{fig3}(b), we use three different paths for the computation of band dispersion in order to capture all high symmetry points in sBZ. An example of strain-induced broken lateral symmetry is depicted in Fig.\ref{fig3}(b) and (c), where we present the electronic bands for strained $\alpha$ superlattice ($\varepsilon_x=-3\%$, $\varepsilon_y=3\%$) along the paths depicted in Fig.\ref{fig3}(b). 

In Fig.\ref{fig3}(d) we present the band dispersion for diverse cases in which the G/hBN superlattices are unstrained, equally strained along the two axes ($x,y$) and subjected to non-equibiaxial strain. Also, note that the strain configurations for which the electronic bands are displayed in the panel (d) are those marked with the black triangle, square and circle in Fig.\ref{fig2}(b-d). Since hBN is a large band gap insulator, most of the contribution to the low energy bands of G/hBN heterostructures is dominated by the pz orbitals of Carbon atoms \cite{Kan2012}. Thus, as discussed earlier, the low energy electronic behavior of G/hBN is mostly governed by that of monolayer graphene \cite{Abergel2013,Wallbank2013, Moon2014}. Therefore, as evident from the first row of Fig.\ref{fig3}(d), the bands of unstrained G/hBN superstructures are linear in ${\vec k}$ and mimic the bands of single layer graphene, however, our computations show that they possess small band gaps ($<$10 meV) which are not clearly observable in Fig.3(d). Furthermore, the band velocity close to the neutrality point is reduced as the twist angle between the layers shrinks. The enhanced substrate-induced renormalization of the Fermi velocity for larger moir\'e patterns yields stronger interactions with the substrate. Overall, the G/hBN superstructures retain the Dirac cones of the single layer graphene with renormalized Fermi velocity.

%
%
%
%
%
\begin{figure}[!t]
\centering
\includegraphics[width= .8\linewidth]{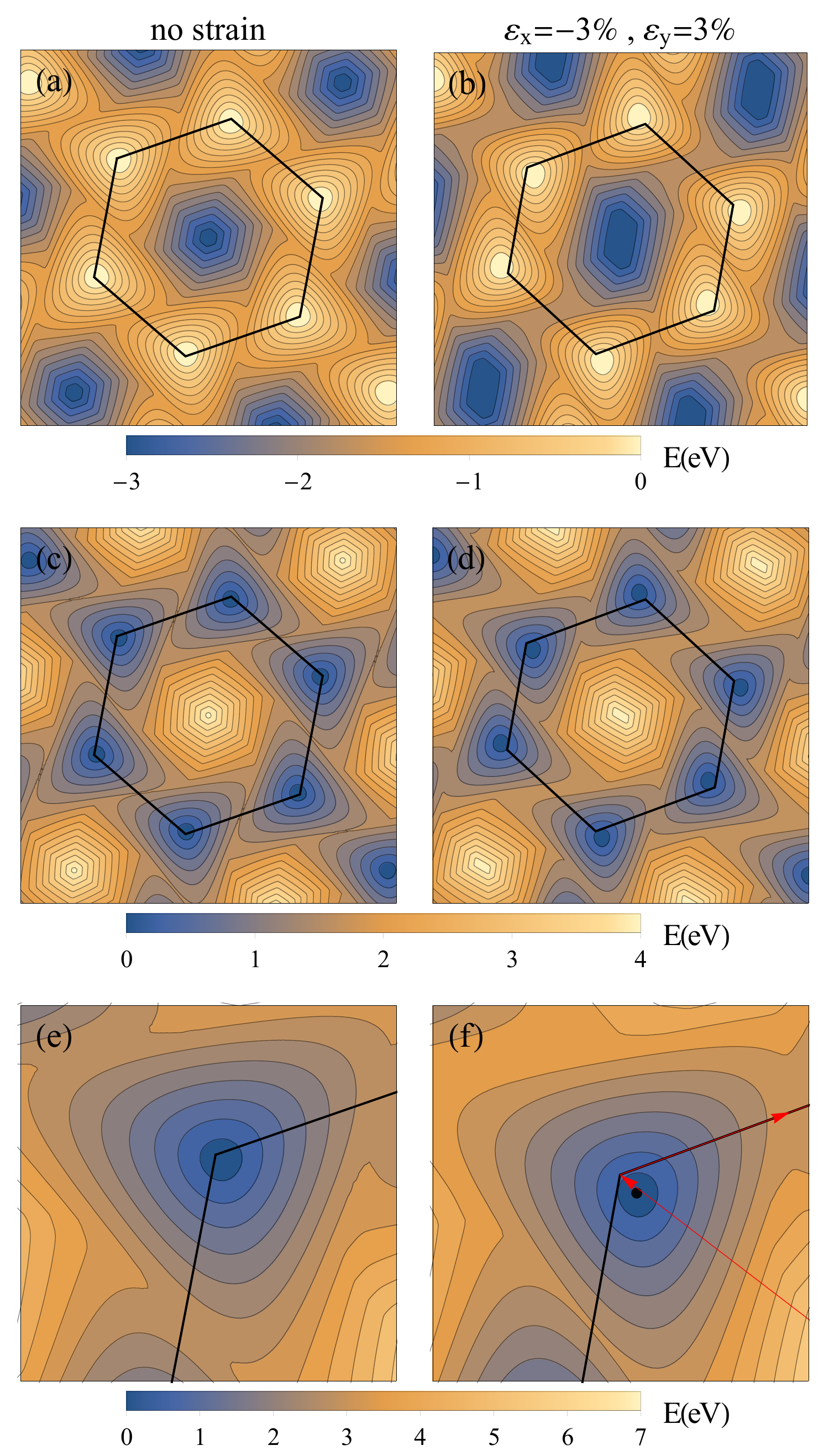}
\caption{(Color online).
Contour maps of (a,b) the valence and (c,d) conduction bands for $\alpha$ superlattice as a function of momentum component in $x$ and $y$ direction. The black hexagons are the sBZ for the corresponding strain configuration. The upper right and lower left corners of the contour plots are $(1,1)×\pi\AA^{-1}$ and $(-1,-1)×\pi\AA^{-1}$ in reciprocal space. The first column corresponds to unstrained $\alpha$ G/hBN and the second column corresponds to the same moir\'e structure with $\varepsilon_x=-3\%$ and $\varepsilon_y=3\%$. (e) and (f) are the energy separation between the valence and conduction bands close to $\kappa_2^{(\prime)}$ for the same strain condition as in previous panels. The upper right and lower left corners of the contour plots are $(-0.15,0.7)×\pi$\AA$^{-1}$ and $(-0.85,0)×\pi$\AA$^{-1}$ in reciprocal space. The filled black circle in the panel (f) is the location of the least band spacing between the conduction and valence bands. The red arrows indicate the path of high symmetry points for which the band dispersion is plotted and displayed in Fig.\ref{fig3}(d). \label{fig4}}
\end{figure}
Biaxial tensile strain preserves the real space symmetries and so does the unstrained band dispersion. As it's shown in the second row of the Fig.\ref{fig3}(d), the renormalized Dirac cones of unstrained G/hBN are preserved in strained corresponding superstructures due to the isotropic expansion of the superlattices when applying biaxial strain. 

The band dispersion for non-equibiaxial strain situations displayed in the third row of Fig.\ref{fig3}(d) is strongly affected by the strain. The strain configurations ($\varepsilon_x$,$\varepsilon_y$) illustrated for the G/hBN superlattices are as follows. (-3\%,3\%) for $\alpha$, (3\%,0) for $\beta$ and (2\%,-3\%) for $\lambda$. While some electronic bands close to one sBZ corner become massive and gapped, the others remain linear and shift away from the sBZ corner. The smallest band spacing between the massive bands is also displaced from the sBZ corners for all G/hBN heterostructures. As an example, it can be seen that the minimum band spacing for strained $\alpha$ superlattice occurs in the vicinity of the $\kappa^\prime_{2}$. Furthermore, the relocation of the valley from the sBZ corners is followed by a discontinuity of band velocity at $\kappa^\prime_{2}$. Similar behavior is observable in the band dispersion of other superstructures, $\beta$, and $\lambda$ when applying non-equibiaxial strain (cf. black circle row of the panel (d)). In biaxially strained G/hBN valley drifts similar to the ones observable for non-equibiaxial strains are absent as a consequence of the fact that the lateral symmetries are preserved (black square row of the panel (d)). 

Our computations show that the band gap energy for the unstrained $\alpha$, $\beta$ and $\lambda$ G/hBN is 4.7, 8.5 and 3.4 meV, respectively. The lateral commensuration strains due to local registry stacking of the G/hBN moir\'e superlattices are responsible for the emergence of band gaps \cite{Jung2017}. We find that the band gaps for biaxially strained ($\varepsilon_x=\varepsilon_y=2\%$) $\alpha$, $\beta$ and $\lambda$ G/hBN shown in the second row of Fig.\ref{fig3}(d), are 4.4, 7.7 and 13.7 meV, severally. Therefore, the band gaps are further enhanced or reduced by biaxial tensile strain depending on the commensuration angle. Non-equibiaxial strains compared to biaxial strains are more efficient in the opening of band gaps for all G/hBN structures.

For a deeper understanding of the valley drift, discontinuity of band velocity and relocation of the gaps we present the contour maps of the highest valence band, lowest conduction band in Fig.\ref{fig4}(a-d) for the $\alpha$ superstructure as an exemplary case. Also, the resolution of the energy separation between the valence and the conduction band is shown in Fig.\ref{fig4}(e,f). Each column is representative of a specific strain configuration. The bands for unstrained $\alpha$ superlattice are demonstrated in the first column and the second column correspond to strained $\alpha$ where $\varepsilon_x=-3\%$ and $\varepsilon_y=3\%$. The isoenergy contours of unstrained and strained G/hBN close to sBZ corners follow the pronounced threefold symmetry of trigonal warping \cite{Ortix2012} which stems from the broken sublattice symmetry of graphene due to the hBN substrate. The trigonal warping of the Dirac fermions in G/hBN superlattice results in anisotropic renormalization of the Fermi velocity. As it is clearly visible from Fig.\ref{fig4}(b) and (d), the Dirac cones of the unstrained G/hBN superstructure is preserved in strained superlattice but with a shift of the conical points to the vicinity of the strained sBZ corners. Consequently, the smallest band spacing also drifts away from the sBZ corner and locates beyond the path of high symmetry points (c.f. the filled black circle in Fig.\ref{fig4}(f)). Therefore, due to the displacement of gaps beyond the high symmetry path, any attempt for evaluation of strain-induced changes of the gap should be made with care.
%
%
%
%
\begin{figure}[!b]
\centering
\includegraphics[width=\linewidth]{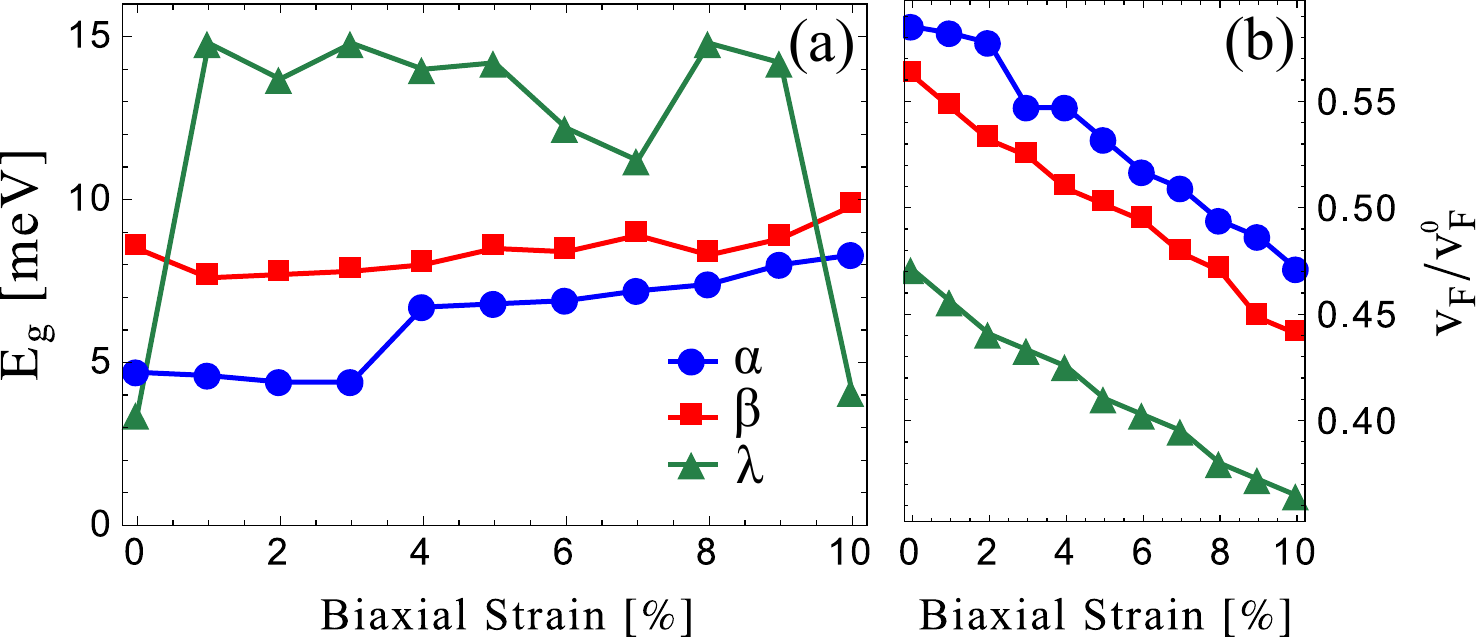}
\caption{(Color online). 
(a) The gap energy and (b) the Fermi velocity as a function of biaxial strain for $\alpha$, $\beta$, and $\lambda$ G/hBN superstructures demonstrated in Fig.\ref{fig1}. $\rm{v^0_F}\approx c/300$ is the monolayer graphene Fermi velocity. \label{fig5}}
\end{figure}

For the strained $\alpha$ G/hBN superstructure shown in Fig.\ref{fig4} we find that the gap energy is $35.1~\rm{meV}$, which in comparison to the unstrained case, ($4.7~\rm{meV}$), is 7 times greater. Therefore, compared to the spontaneous strains which only have strong effects on DOS and can weakly affect the band gap \cite{San-Jose2014}, gap considerably increases for special cases of in-plane tensile strain even though the strain itself is small ($3\%$ for this case). Also, compared to the case of mono and bilayer graphene where spectral gaps emerge for strain magnitudes greater than 20\% and only along the preferred direction \cite{Pereira2009a,Verberck2012}, moir\'e structures are more efficient in gap opening and can considerably enhance the strain impacts.

Fig.\ref{fig5}(a) illustrates the gap energy as a function of biaxial strain for $\alpha$, $\beta$ and $\lambda$ superstructures. Similar to the case of non-equibiaxial strain presented in Fig.\ref{fig4}, the equibiaxial strain leads to the modification of the gap energy except that the strain-induced changes of the hybridization of the atomic bonds of G/hBN are isotropic leading to relatively smaller changes of the gap energy. Despite the resembling behavior of the strain energy for all superlattices (Fig.\ref{fig2}), the modulation of the band gap energy is different for each superlattice when applying biaxial tensile strain. Thus, contrary to the strain energy which is not dependent on $\theta_{m,n}$, the gap energy is strongly affected by the misalignment angle and the stacking of the graphene layer on hBN (see Fig.\ref{fig3}). Except for $\lambda$, all superlattices show non-vanishing and relatively small gap energy not exceeding $10~\rm{meV}$. Moreover, the gap energy increases monotonically with increasing strain for $\alpha$ and $\beta$ superlattices. The $\lambda$ superstructure has the largest gap excluding the situations in which the applied strain is $0\%$ and $10\%$. Overall, since strain is a costly method in tunning and opening of the bandgaps for monolayer graphene \cite{Pereira2009a}, the G/hBN heterostructures are intriguing platforms for tuning the electronic structure of graphene at low strain costs, specifically because the fabrication of the moir\'e superstructures with controlled alignment of layers is experimentally practical \cite{Kim2016,Yankowitz2016a}. 

We next calculate the Fermi velocity, $\rm{v_F}=\frac{1}{\hbar}\frac{\partial E}{\partial k}$, by fitting a linear dispersion to the bands close to the conical point. The band velocity close to the neutrality point for all superstructures is depicted in Fig.\ref{fig5}(b). In the unstrained G/hBN superstructures, the Fermi velocity is reduced and renormalized to almost half of the single layer graphene Fermi velocity which is in good agreement with previously reported values \cite{Yankowitz2012,Ortix2012}. Moreover, the substrate-induced renormalization of the Fermi velocity scales inversely with the misalignment angle yielding that the interlayer interactions are stronger for larger moir\'e patterns. Furthermore, the Fermi velocity is suppressed when applying biaxial tensile strain. We find that the strain-induced rate of the Fermi velocity reduction is approximately -0.011 $\rm{v^0_F}$ per percent of applied biaxial strain for all three G/hBN superstructures displayed in Fig.\ref{fig5}. Since the isotropic expansion of the lattice does not alter the stacking-dependent symmetries, the strain-induced changes of the Fermi velocity are equal for all investigated G/hBN superstructures.

%
%
%
%
\begin{figure}[!t]
\centering
\includegraphics[width=.8\linewidth]{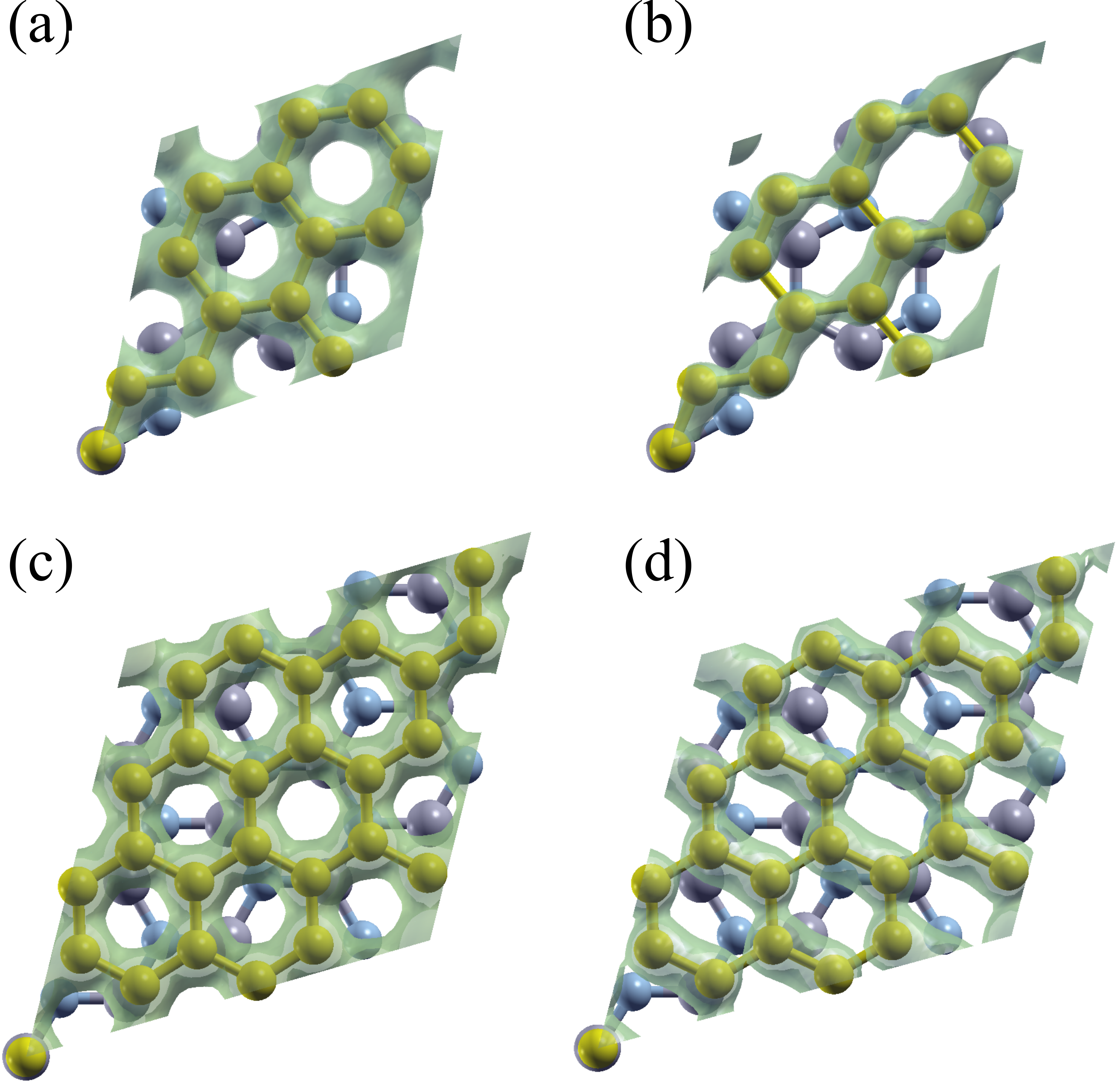}
\caption{(Color online). Local density of states (LDOS) for two exemplary strain configuration on (a,b) $\alpha$ and (c,d) $\beta$ G/hBN superlattices. The energy interval over which the LDOS is calculated is [-0.2,0] eV and belongs to the highest valence band presented in Fig.\ref{fig3}. Panel (a) and (c) depict the charge density for the unstrained G/hBN superlattices and panel (b) and (d) are for those that are under non-equibiaxial strain, i.e. $\varepsilon_x=3\%$ and $\varepsilon_y=-3\%$. Isovalue for all configurations is set to $4.5\times10^{-5}~e/\AA^3$.
\label{fig6}}
\end{figure}
Fig.\ref{fig6} is the resolution of the charge density over the energy interval of [-0.2,0] eV for unstrained and strained (a,b) $\alpha$ and (c,d) $\beta$ G/hBN superstructures. When strained, $\alpha$ and $\beta$ G/hBN superlattices are 3\% stretched along the $x$ axis and 3\% compressed along the $y$ axis. Both superlattices demonstrate a homogeneous spatial distribution of the charge density over the graphene layer when the applied strain is zero (See Fig.\ref{fig6}(a) and (c)). Due to large misalignment angle, the weak interlayer interaction between graphene and hBN layers leaves the layers electronically decoupled leading to weak sublattice symmetry breaking and almost equal expansion of the charge density over different carbon sublattices. On the other hand, we interestingly observe that the charge density is aligned along the zigzag direction when both superstructures are exposed to mixed strain configuration $(\varepsilon_x,\varepsilon_y)=(3\%,-3\%)$. The zigzag direction remains more efficient for electronic transport whilst having a different alignment in the investigated G/hBN superlattices of $\alpha$ and $\beta$ and also compared to the imposed strain alignment. Thus, similar to the monolayer graphene \cite{Pereira2009a}, strain-induced anisotropic transport becomes one dimensional but at low strain cost and by suppression of bonds when applying in-plane non-equibiaxial strain.
\section{Conclusion}\label{concl}
We have studied the impacts of the in-plane strain on the electronic behavior of commensurate G/hBN superlattices. We found that the substrate-induced trigonal warping of the Dirac fermions and the Fermi velocity renormalization are dependent on the in-plane strain. Also, the asymmetric strains cause valley drifts and relocation of the band gap from the sBZ corners and the path of the high symmetry points. Therefore, the identification of the true band gap becomes non-trivial and should be done with care. We found that the Fermi velocity decreases when applying biaxial tensile strain. Furthermore, large direct gap energy emerges for small non-biaxial in-plane strain imposition ($3\%$). Thus, compared to the case of mono and bilayer graphene where spectral gaps emerge for strain magnitudes greater than 20\%, G/hBN moir\'e superstructures can considerably enhance and magnify the strain impacts. This implies that the G/hBN heterostructures are interesting platforms for tuning the electronic structure of graphene at low strain costs which can be exploited in graphene-based nanoelectronic devices such as vertical tunneling  transistors \cite{Ferrari2014}.
\section{Acknowledgments}
Authors acknowledge the support by the Cluster of the school of nano science at the Institute for Research in Fundamental Sciences (IPM).
\bibliography{ref.bib}
\end{document}